\documentclass[aps,prl,twocolumn,showpacs]{revtex4}
\usepackage[T1]{fontenc}
\usepackage[latin1]{inputenc}
\usepackage[dvips]{graphicx}
\usepackage{hyperref}
\usepackage{amsmath}
\usepackage{ulem}



\begin{document}
\title{Anomalous conductivity: impact of nonlinearity and disorder}
\author{M.~V.~Ivanchenko$^{1,2}$ and S.~Flach$^2$}

\affiliation{$^1$ Theory of Oscillations Department, University of Nizhniy Novgorod, Russia
\\ $^2$ Max-Planck-Institut f\"ur Physik komplexer Systeme, N\"othnitzer Str. 38, 01187 Dresden, Germany}

\begin{abstract}
We reveal the intricate impact of nonlinearity and disorder on the thermal conductivity of acoustic chains. 
Disorder induces mobility edges and allows to control the amount of extended modes which are the ballistic channels for energy transport.
Nonlinearity adds a diffusive conductivity channel through interacting localized modes, and controls the contact resistance at the edges.
Analytical arguments and numerical results yield several crossovers between dominating heat channels,
when varying the system size and the temperature. 
We demonstrate that the nonlinearity-induced interaction between the modes alters the transport through existing conductivity channels and creates new ones, 
underpinning the observed phenomena.
\end{abstract}


\maketitle

Deviation from the normal (Fourier-law) heat conductivity has recently become a matter of experimental and applied interest in the context of nanotube systems, inspiring the research on solid-state thermal rectifiers and nanotube phonon waveguides \cite{devices}. The anomalous conductivity was clearly manifested in the pioneering experiments with carbon and boron-nitrid nanotubes, yielding $\mathcal{K}\propto N^\alpha, \ \alpha\sim0.5 - 0.6$ \cite{Chang}, with the other promise coming from nanowires \cite{nanowires}. In both systems phonons were reported to be the main heat flux carriers. Phonon scattering and interaction due to disorder (impurities) and nonlinearity (atomic interactions) are, thus, likely to determine the anomaly and its temperature dependence. Simulations with realistic models indicate that disorder \cite{Stoltz} or nonlinearity \cite{savin} can give anomalous heat conductivity, but the 
results on the exponent $\alpha$ are conflicting.

One-dimensional momentum-conserving arrays serve as simple qualitative models to study
anomalous heat conduction \cite{early_results}. Despite intense research and some qualitative understanding, we still lack quantitative agreement on
the main characteristics of anomalous conductivity \cite{Lepri_review,Matsuda,predictions}. 
Moreover, mostly harmonic chains with disorder, or anharmonic ordered chains were studied.
Harmonic systems do not equilibrate and the conductivity depends on the boundary conditions and the spectrum of thermal noise. For anharmonic ordered systems 
one lacks control over the number of relevant long wavelength modes which contribute to anomalous conductivity.
The interplay between disorder and anharmonicity was touched in Refs. \cite{BambiHu,Dhar}, where the regime of normal conductivity \cite{BambiHu} was questioned
due to finite size effects \cite{Dhar}.

In this Letter we uncover and study the intricate impact of nonlinearity on the thermal conductivity of the disordered 
Fermi-Pasta-Ulam (FPU) chain with fixed boundaries. Upon variation of the temperature and the chain size we observe transitions between the following regimes: (i) insulating behavior with $\alpha<0$, (ii) normal-like conductivity, $\alpha\sim 0$, (iii) disorder-dominated anomalous conductivity, $\alpha\sim 0.52 - 0.58$, (iv) nonlinearity-dominated anomalous conductivity, $\alpha\sim 0.38$. 
The crossovers and the scaling of $\mathcal{K}$ are explained by analyzing the properties and interaction of localized and delocalized modes, which are continued into the nonlinear regime, and emerging new heat conductivity channels.
We show that {\it anharmonic disordered} systems offer a better way to study the mechanisms of anomalous conductivity, being also more realistic models for experimental setups.

We consider the FPU-$\beta$ chain of $N$ equal masses, with disorder in the harmonic spring constants, and additional quartic anharmonicities in the spring potential:
\begin{equation}
\label{eq1}
\begin{aligned}
& H=\frac{1}{2}\sum\limits_{n=1}^{N} p_{n}^2
+\sum\limits_{n=1}^{N+1}\left[\frac{1}{2}(1+D\kappa_{n})(x_{n}-x_{n-1})^2\right.\\
&\left.+\frac{\beta}{4}(x_{n}-x_{n-1})^4 \right] 
\end{aligned}
\end{equation}
where $x_{n}(t)$ is the displacement of the
$n$-th particle from equilibrium, $p_{n}(t)$ its momentum,
$\kappa_{n}\in[-1/2,1/2]$ are random, uniform, and uncorrelated, $\left\langle \kappa_{n}\kappa_{m}\right\rangle=\sigma^2_{\kappa}\delta_{n,m}$. We apply fixed boundary conditions $x_0=x_{N+1}=0$.

The heat conductivity is measured by a standard approach, when the thermal baths attached to the ends generate a temperature gradient and heat current  along the chain \cite{Lepri_review}.  We implement Nos\'e-Hoover thermostats adding the terms $-\zeta_\pm\dot{x}_{1,N}$ to the respective equations of motion, where $\dot{\zeta}_\pm=\dot{x}_{1,N}^2/T_\pm-1$. 
The heat flux along the chain is defined as the time average of $j=-\frac{1}{2}\sum\limits_n(\dot{x}_{n+1}+\dot{x}_n)[(1+D\kappa_{n+1})(x_{n+1}-x_{n})+\beta(x_{n+1}-x_{n})^3]$
\cite{Lepri_review}. The heat conductivity coefficient reads $\mathcal{K}=j N/(T_+-T_-)$ then. We make use of the mean temperature $T=(T_++T_-)/2$  as a parameter corresponding to the energy density 
$\langle E_n\rangle=k_BT=T$, setting $k_B=1$ and $(T_+-T_-)/T=0.5$ further on.

We start with the canonical transformation to and from the normal modes of the harmonic lattice in the absence of disorder:
$x_{n}(t)=\sum_{q=1}^N Q_{q}(t)
z_{qn}$ which defines the mode space with $N$ coordinates
$Q_{q}(t)$ and eigenvectors $z_{qn}=\sqrt{\frac{2}{N+1}}\sin{\left(\frac{\pi q n}{N+1}\right)}$. In the presence of anharmonicity and disorder
the dynamics of the modes follows
\begin{equation}
\label{eq2}
\begin{aligned}
 &\Ddot{Q}_{q}+\omega_{q}^2
 Q_{q}=-\nu\sum\limits_{p,r,s=1}^N
 C_{q,p,r,s}\omega_q\omega_p\omega_r\omega_s Q_{p}Q_{r}Q_{s}\\
 &-d\sum\limits_{p=1}^N \omega_q\omega_p K_{q,p} Q_p\;.
 \end{aligned}
\end{equation}
Here $\omega_{q}=2\sin{\frac{\pi q}{2(N+1)}}$ are the normal mode frequencies. 
The coupling coefficients
$C_{q,p,r,s}$ \cite{we_qb}
control the selective anharmonic interaction between
modes, and the coefficients $K_{q,p}=\frac{2}{N+1}\sum_{n=1}^{N} \kappa_n \cos \frac{\pi q (n-1)/2}{N+1} \cos \frac{\pi p (n-1)/2}{N+1} $ \cite{qb_do} reflect the all-to-all 
linear interaction due to disorder. The nonlinearity and disorder parameters $\nu=\beta/(N+1),\,d=D/\sqrt{N+1}$ are small: $\nu, d\ll 1$ for $\beta=D=1$ and large system
size $N \gg 1$.

Let us consider the case $d\neq 0$ and $\nu=0$. We aim at computing the new eigenvectors $\hat{z}_{qn}$ defined through the transformation
$x_{n}(t)=\sum_{q=1}^N Q_{q}(t) \hat{z}_{qn}$.
We apply a perturbational approach for the harmonic mode $q_0$ using the small disorder parameter $d$: ${Q}_q(t)={Q}_q^{(0)}(t)+d{Q}_q^{(1)}(t)+\ldots$, where $Q^{(0)}_q(t)=0$ for $q\neq q_0$. In the first order approximation, from Eq. (\ref{eq2}) we obtain the equation of a forced oscillator for modes with $q\neq q_0$:
$\Ddot{Q}_{q}^{(1)}+\omega_{q}^2 Q_{q}^{(1)}=-\omega_q\omega_{q_0} K_{q,q_0} Q_{q_0}^{(0)}$. As a result we find for the amplitude $A$ of each mode
\begin{equation}
\label{eq2b}
\begin{aligned}
 A_{q,q_0}^{(1)}=-\frac{\omega_q\omega_{q_0}}{\omega_q^2-\omega_{q_0}^2} K_{q,q_0} A_{q_0}, \ q\neq q_0
 \end{aligned}
\end{equation} 
Accordingly, the time-averaged $q$-th mode energy is
\begin{equation}
\label{eq2c}
\begin{aligned}
\left\langle E_{q}\right\rangle=\frac{d^2 E_{q_0}\omega_q^4}{2(\omega_q^2-\omega_{q_0}^2)^2}\left\langle K_{q,q_0}^2\right\rangle=\frac{d^2 E_{q_0} \sigma_\kappa^2\omega_q^4}{2(\omega_q^2-\omega_{q_0}^2)^2}\;.
 \end{aligned}
\end{equation}
For Eq. (\ref{eq2b}) being valid we request $\left\langle E_{q_0+1}\right\rangle\ll E_{q_0}$, hence, 
\begin{equation}
\label{eq6}
\begin{aligned}
q_0\ll q_c=2\sqrt{2}(N+1)^{1/2}/ D \sigma_\kappa\;.
 \end{aligned}
\end{equation}

It follows that normal modes with mode numbers $ q_0 \ll q_c$ approximately keep their plane wave eigenvector profile 
$\hat{z}_{q_0n}=\sqrt{\frac{2}{N+1}}\left(\sin\frac{\pi q_0 n}{N+1}-d \sum\limits_{p\neq q_0} \frac{\omega_{q_0}\omega_{p}}{\omega_{q_0}^2-\omega_{p}^2} K_{q_0,p}\sin\frac{\pi p n}{N+1}\right)$ in real space in the presence of disorder. Therefore these {\sl metallic} modes
are still delocalized in real space. On the other side, Anderson localization implies that the eigenmodes of a one-dimensional disordered chain are localized.
Therefore $q_c$ sets a mobility edge: for $q_0 \ll q_c$ the metallic eigenmodes are delocalized in real space, and for 
$q_0 > q_c$ the {\sl insulating} eigenmodes are localized. Note, that in short chains or/and at weak disorder {\it all modes} are metallic and delocalized if $q_c\ge N$, (\ref{eq6}) yielding $N\le 8/D^2\sigma_{\kappa}^2$.
Remarkably, the transfer matrix approach 
yields a lower boundary $q_c$ for spatially localized eigenstates (computing their localization length) \cite{Matsuda} with the 
same scaling $q_c\propto N^{1/2}$. 
These results tell, that for the disordered harmonic chain there is a thin layer of $\sqrt{N}$ metallic modes in frequency space with frequencies
$0 \leq \omega \leq \omega_c \sim N^{-1/2}$. For most practical purposes these modes can be described using the eigenvectors of the ordered harmonic chain.

In a next step we add anharmonic terms, and compute the nonlinear analogs of normal modes - q-breathers \cite{we_qb} for the metallic modes of the disordered harmonic chain.
We introduce $\{\hat{Q}_q,\hat{P}_q\}$ such that $x_{n}(t)=\sum\limits_{q=1}^N \hat{Q}_{q}(t) \hat{z}_{qn}$. It follows:
\begin{equation}
\label{eq7}
\begin{aligned}
 &\Ddot{\hat{Q}}_{q}+\omega_{q}^2
 \hat{Q}_{q}\approx -\frac{\nu}{2}\sum\limits_{p,r,s=1}^{q_s}
 C_{q,p,r,s}\omega_q\omega_p\omega_r\omega_s \hat{Q}_{p}\hat{Q}_{r}\hat{Q}_{s}\;.
 \end{aligned}
\end{equation}

We develop the perturbation theory in powers of $\nu$, taking $\hat{z}_{q_0n}$ as the zero order of approximation. Neglecting the higher order terms $\mathcal{O}(\nu d)$, we obtain the q-breather \cite{we_qb} with an energy distribution exponentially localized in the metallic mode space:
\begin{equation}
\label{eq4}
E_{(2n+1)q_0}=\lambda^{2n}E_{q_0}\;,\;
\lambda=\frac{3\beta
E_{q_0}(N+1)}{8\pi^2 q_0^2}\;\;
\end{equation}
About the mobility edge the modes get extremely weak energies $E_{q_c}\sim\lambda^{q_c/q_0}E_{q_0}$ once $\lambda<1$ and the excitation of insulating modes is negligible.

More than that, one can construct $s$-dimensional q-tori by continuing the first $s\ll q_c$ disordered modes into the nonlinear regime, as done in the ordered case \cite{tassos}. These objects will be exponentially localized in q-space:
\begin{equation}
\label{eq9a}
E_{p}/E_0\propto(\lambda_T)^{2p}\;,\;
\lambda_T=\beta N E_0/s\;\;
\end{equation}
where $E_0$ and $E_p$ are the average energies of the modes $\overline{1,s}$ and $\overline{(2p-1)s+1,(2p+1)s}$ \cite{tassos}. Delocalization of tori happens when $\lambda_T>1$. Taking $s\propto q_c$ we obtain the following strong stochasticity threshold:
\begin{equation}
\label{eq9b}
\begin{aligned}
N_c\propto1/D^2\sigma_\kappa^2 \beta^2 E_0^2\;.
\end{aligned}
\end{equation}

\begin{figure}[t]
{\centering
\resizebox*{0.95\columnwidth}{!}{\includegraphics{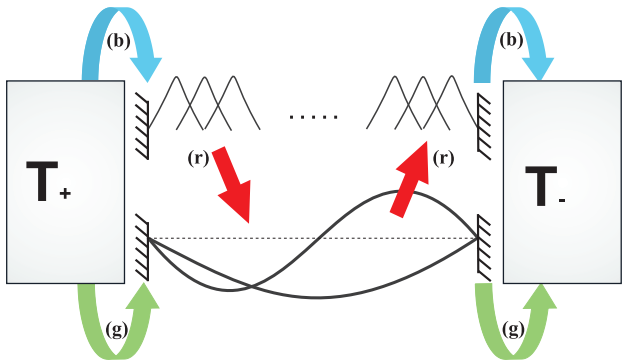}}}
{\caption{(Color online) Schematic representation of conductance channels. Thick black metallic modes and thin black insulating modes are shown. (B)lue and (g)reen arrows show the energy fed into localized and delocalized modes channels by direct interaction  with the heat baths. Both carry the heat fluxes (the latter more efficiently). Nonlinearity induces the heat flux via the localized modes into delocalized ones ((r)ed arrows), the third channel. 
}\label{fig0}}
\end{figure}

Below the strong stochasticity threshold one can think of three basic heat conductivity channels (Fig.\ref{fig0}). (I) Metallic modes carry the heat flux  ballistically by direct interaction with the heat baths, as in the linear case. The magnitude of interaction and the mode-specific heat flux are proportional to the square amplitudes next to the boundaries $j_q\propto \hat{z}_{q,n=1,N}^2$ (boundary resistance) \cite{Lepri_review}. For fixed ends $j=\sum_{q<q_c}j_q\propto N^{-3/2}$ and $\mathcal{K}\propto N^{-1/2}$. (II) Insulating localized Anderson modes 
start to interact with each other due to anharmonicity, and are expected to contribute a normal diffusive energy transport with $\mathcal{K}(T)$ independent of $N$ \cite{onsite}.
(III) Anharmonicity also opens the third conductivity channel, by decreasing the boundary resistance for metallic modes.
This happens due to thermalization among insulating modes, which in turn will interact with the metallic modes in a broad range of space, thus eliminating the effects of
fixed boundary conditions.

The first conductivity channel (I) can be dominant only when the mean temperature $T$ is low and the system length $N$ is small enough. 
The heat current via the second localized mode channel (II) can be estimated using recent studies of wave packet evolutions in disordered Klein-Gordon chains,
where the harmonic approximation yields an Anderson insulator with a finite upper bound on the localization length of each eigenmode \cite{spreading}. 
For a given energy density (respectively finite temperature) the diffusion rate (and therefore also the heat conductivity) are predicted to
vary as $\mathcal{K} \sim T^2$ for $T > T_{cr}$ and $\mathcal{K} \sim T^4$ for $T < T_{cr}$. Here the crossover temperature $T_{cr}$ is a model-dependent
temperature which depends on the strength of disorder. Notably this heat conductivity does not depend on the size of the system.
Therefore the second channel will dominate over the first one if $N > N_{12}(T)$ where $N_{12}(T)$ is some function which increases with decreasing temperature.
The crossover to normal conductivity will be observed at $N\approx N_{12}(T)$.

Similarly, the heat flux through the third conductivity channel can be estimated. Assume that $\mathcal{O}(N)$ strongly localized insulating
modes $\xi_{q}=\mathcal{O}(1)$ are distributed uniformly along the chain, $E_q=T$, and heat the delocalized metallic mode $q_0$ with initial energy $E_{q_0}=0$. It follows $\ddot{Q}_{q_0}+\omega_{q_0}^2 Q_{q_0}\approx\beta \sum_{p>q_c} I_{q_0,p} T^{3/2} \zeta(t)$ with 
$\zeta(t)$ being an uncorrelated Gaussian noise with unit variance.
Then $\langle \dot{Q}_{q_0}^2\rangle\propto \beta^2 \omega_{q_0}^2 T^3 t$ and the heat flux into the delocalized mode is $J_{q_0}\propto \beta^2 \omega_{q_0}^2 T^3$. Extrapolating this result to the equilibrium conductivity problem, one obtains $J=\sum_{q_0<q_c}J_{q_0}\propto \beta^2 T^3 N^{-1/2}$ and the associated heat conductivity $\mathcal{K}\propto N^{1/2}$. 
Thus, for $N > N_{23}(T)$ we expect to observe a crossover from normal conductivity to anomalous $\alpha\approx 0.5$. 
Finally, with further increasing of the system size $N$, the strong stochasticity threshold (\ref{eq9b}), $E_0\equiv T$, is reached for $N > N_{sst}$. 
Then metallic modes start to strongly interact with each other, and theoretical predictions of fully developed turbulence within renormalization group approaches,
and mode coupling theories may become applicable with the consequence that one will observe $\alpha\sim 1/3\ldots 2/5$.

To summarize, we expect that depending on the systems size, heat is carried by:
(i) ballistic metallic modes directly coupled to the heat reservoirs for $N < N_{12}$ with $\alpha \approx -1/2$; (ii) insulating localized modes interacting with each other for $N_{12} < N < N_{23}$ with $\alpha \approx 0$;
(iii) ballistic metallic modes which are coupled to the heat reservoir via insulating localized modes for $N_{23} < N < N_{sst}$ with $\alpha \approx  1/2$;
(iv) strongly interacting metallic modes which are coupled to the heat reservoir via insulating localized modes for $N_{sst} < N$ with $\alpha \approx 1/3 ... 2/5$ .

\begin{figure}[t]
{\centering
\resizebox*{0.95\columnwidth}{!}{\includegraphics{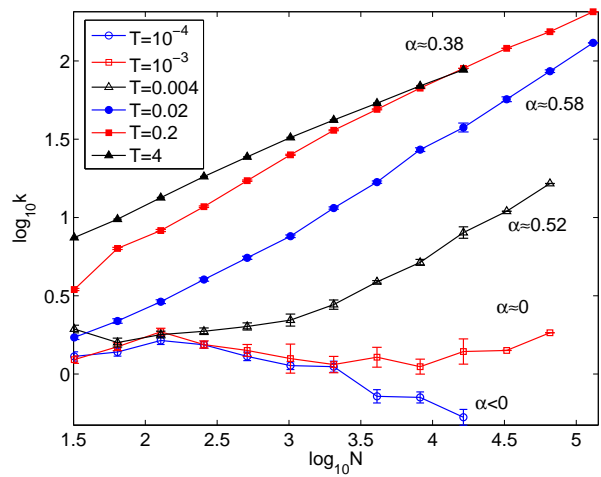}}}
{\caption{(Color online) Averaged heat conductivity coefficient vs. chain length in the log-log scale at different mean temperatures and $\beta=D=1$. The values of slope estimated by the linear fit are printed. Error bars estimate statistical error.}\label{fig1}}
\end{figure}

\begin{figure}[t]
{\centering
\resizebox*{0.95\columnwidth}{!}{\includegraphics{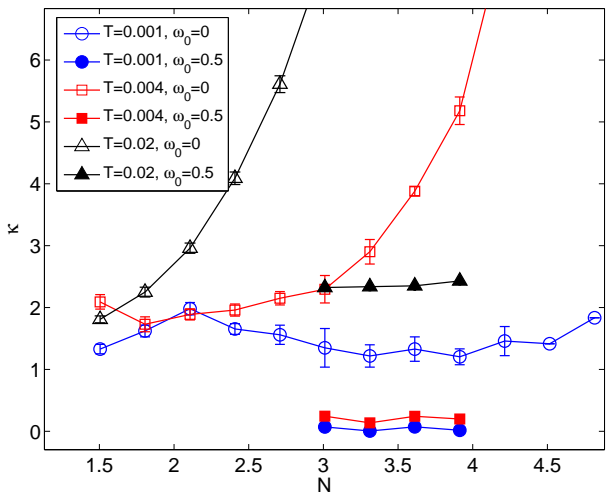}}}
{\caption{(Color online) Size dependence of the heat conductivity without and with additional random on-site harmonic potential (open and filled markers respectively).}\label{fig2}}
\end{figure}

Let us turn to the numerical results, obtained with the integration time varying from $10^6$ for the shortest chains to $2\cdot10^7$ for the longest, and up to 100 disorder
realizations. In all simulations $\beta=D=1$. As observing all crossovers at a single temperature is currently impossible (the required chain sizes and integration times are unattainable), we 
gradually increase $T$. The results need careful interpretation, as the finite size and residual temporal transient effects (even after long equilibration times) cannot be determined at present.

At $T=10^{-4}$ we observe the first conductivity channel I with $\alpha<0$  (Fig.\ref{fig1}). For larger temperature $T=0.001$ 
we observe the normal-like conductivity of regime II in the whole range of studied $N$, indicating that the localized modes conductivity channel is dominant. At $T=0.004$ we observe the crossover to the anomalous conductivity channel III with the value $\alpha\approx0.52$ for large $N$, with even slightly bigger $\alpha$ at higher temperatures. It corresponds to the ballistic transport by metallic modes coupled to the heat baths via insulating localized modes. 

To confirm that this anomaly is due to the metallic ballistic modes, we add a random on-site harmonic potential $\frac{1}{2}\sum_{n=1}^N\Omega_n^2x_n^2$, $\Omega_n\in [0;\omega_0]$ to (\ref{eq1}).
Then a gap around zero frequencies opens in the frequency spectrum of the harmonic system, total momentum is not anymore conserved, and all modes become insulating and localized. Indeed, numerical results demonstrate the absence of anomalous behavior  (Fig.\ref{fig2}). The normal conductivity channel is affected too (due to increased localization of modes), but is still observable and becomes dominating.

Finally, we observe the crossover due to the strong stochasticity threshold which yields  $\alpha\approx0.38$ ($T=0.2$, Fig.\ref{fig1}), in which the temperature dependence of $\mathcal{K}$ becomes indistinguishable. As all other crossovers, 
it can be observed for smaller values of $N$ with increasing $T$ and at $T=4$ this regime dominates starting from the smallest system sizes studied. 
Hence, the crossover $N_c(T)$ can be estimated as the intersection point of the linear fit for $\mathcal{K}(N)$ at $T=4$ and the linear fit of $\mathcal{K}(N)$ at a lower temperature, in its part where $\alpha\sim 0.5$ is observed. For $T=0.02$ and $T=0.2$ one gets $N_c(0.02)\approx5.8\cdot10^5$ and $N_c(0.2)\approx4.5\cdot10^3$, which agrees with $N_c\propto T^{-2}$ (\ref{eq9b}) well.

In conclusion, we shed light on the complex interplay between disorder and nonlinearity that determines the heat conductivity of acoustic chains, revealing a much more intricate picture than suggested before. Depending on the system size and temperature, one can find different conductivity channels to be dominant and observe crossovers as the parameters are changed. The mechanisms are: (i) ballistic transfer by metallic delocalized modes coupled directly to the heat baths, (ii) diffusive transfer by the insulating localized modes, (iii) 
ballistic transfer by metallic delocalized modes, the heat flux coming indirectly from the heat baths mediated via the insulating localized modes,
(iv) turbulent transfer by metallic delocalized modes above the stochasticity threshold. The corresponding size dependence of the conductivity coefficient is drastically different: insulating, normal, and two kinds of anomalous divergence. The studied system size is comparable to the number of atoms along nanotubes; therefore, the predicted crossovers may prove to be observable even in the current experimental systems, if the temperature is varied.

Computations were carried out at the HPC University of Nizhniy Novgorod and MPIPKS. The support of Prof. V.P. Gergel is appreciated. 
We thank 
G. Casati, S. Lepri, R. Schilling, S. Aubry, N. Li for helpful discussions.


\begin{thebibliography}{99}
\bibitem{devices} C.W. Chang {\it et al.}, Science {\bf 314}, 1121 (2006); C.W. Chang {\it et al.}, Phys. Rev. Lett. {\bf 99}, 045901 (2007).
\bibitem{Chang} C.W.~Chang {\it et al.}, Phys. Rev. Lett. {\bf 101}, 075903 2008; 
\bibitem{nanowires} K. Schwab {\it et al.}, Nature {\bf 404}, 974 (2000).
\bibitem{Stoltz} G.~Stoltz, M.~Lazzeri, and F.~Mauri, J. Phys.: Condens. Matter {\bf 21}, 245302 (2009).
\bibitem{savin} A.~V.~Savin, B.~Hu, and Y.~S.~Kivshar, Phys. Rev. B {\bf 80}, 195423 (2009).
\bibitem{early_results} D.N. Payton, M. Rich, W.M. Visscher, Phys. Rev. {\bf 160}, 706 (1967); E.A. Jackson, J.R. Pasta, J.F. Waters, J. Comput. Phys. {\bf 2}, 207 (1968); K.~Ishii, Suppl. Prog. Theor. Phys., {\bf 53}, 77 (1973); T. Prosen and D.K. Campbell, Phys. Rev. Lett. {\bf 84}, 2857 (2000).
\bibitem{Lepri_review} S. Lepri, R. Livi, and A. Politi, Phys. Rep. {\bf 377}, 1 (2003); 
A. Dhar, Adv. Phys. {\bf 57} 457 (2008).
\bibitem{Matsuda} H. Matsuda, K. Ishii, Suppl. Prog. Theor. Phys. {\bf 45}, 56 (1970).
\bibitem{predictions} O. Narayan, S. Ramaswamy, Phys. Rev. Lett. {\bf 89}, 200601 (2002); S. Lepri, Phys. Rev. E {\bf 58}, 7165, (1998);
 A. Dhar, Phys. Rev. Lett. {\bf 86} 3554 (2001); G. Casati, T. Prosen, Phys. Rev. E {\bf 67}, 015203 (2003); G.R. Lee-Dadswell, B.G. Nickel and C.G. Gray, Phys. Rev. E {\bf 72}, 031202 (2005).
\bibitem{BambiHu} B. Li, H. Zhao, and B. Hu, Phys. Rev. Lett. {\bf 86}, 63 (2001).
\bibitem{Dhar} A.~Dhar and K.~Saito, Phys. Rev. E {\bf 78}, 061136 (2008).
\bibitem{we_qb} S.~Flach, M.~V.~Ivanchenko and O.~I.~Kanakov,
Phys. Rev. Lett. {\bf 95}, 064102 (2005); S.~Flach, M.~V.~Ivanchenko and O.~I.~Kanakov, Phys. Rev. E {\bf 73}, 036618 (2006).
\bibitem{qb_do} M.~V.~Ivanchenko, Phys. Rev. Lett. {\bf 102}, 175507 (2009).
\bibitem{tassos} H.~Christodoulidi, C.~Efthymiopoulos, and T.~Bountis, Phys. Rev. E {\bf 81}, 016210 (2010).
\bibitem{onsite} G. Casati et al., Phys. Rev. Lett. 52, 1861 (1984); B. Hu, B. Li, and H. Zhao, Phys. Rev. E 61, 3828 (2000). 
\bibitem{spreading} S. Flach et al, Phys. Rev. Lett. 102, 024101 (2009); Ch. Skokos et al, Phys. Rev. E 79, 056211 (2009); Ch. Skokos et al, Phys. Rev. E 82, 016208 (2010);
T. V. Lapteva et al, EPL 91, 30001 (2010).

\end{thebibliography}
\end{document}